\documentclass[a4paper,twoside,11pt]{article}
\usepackage[utf8]{inputenc}

\usepackage[english]{babel}
\usepackage{graphicx}
\usepackage{subeqnarray}

\usepackage{array}
\usepackage{xcolor}
\usepackage{amsmath}
\usepackage{amssymb}
\usepackage[final]{pdfpages}
\usepackage{graphicx}
\usepackage{bbold}
\usepackage{dsfont}
\usepackage{pgfplots}
\usepackage{subfig}
\usepackage{algorithm}
\usepackage{algorithmic}
\usepackage[font=small,labelfont=bf]{caption}
\usepackage{multirow}

\usepackage[font={footnotesize}]{caption}
\usepackage[utf8]{inputenc}
\usepackage[T1]{fontenc}
\usepackage{xcolor,amsmath,amssymb}
\usepackage{geometry}
\usepackage{graphicx}
\usepackage[sort&compress,numbers]{natbib}
\usepackage{enumerate}
\usepackage{wasysym}
\usepackage[textsize=footnotesize]{todonotes}
\usepackage{url}
\usepackage{pgfplots}

\definecolor{darkblue}{rgb}{0,0,0.6}
\definecolor{darkred}{rgb}{0.6,0,0}

\usepackage[colorlinks=true, urlcolor=black, citecolor=black, linkcolor=black, hyperfootnotes=false]{hyperref}

\usepackage{authblk}

\usepackage[bitstream-charter]{mathdesign}

\usepackage{lipsum}
\graphicspath{{figures/}} 

\DeclareMathOperator{\Cov}{Cov}
\DeclareMathOperator{\Cor}{Cor}
\DeclareMathOperator{\Tr}{Tr}
\newcommand{\comment}[1]{}

\title{Non-parametric Estimation of Quadratic Hawkes Processes\\ for Order Book Events}

\author[1,2]{Antoine Fosset}
\author[2,3]{Jean-Philippe Bouchaud}
\author[1,2,3]{Michael Benzaquen\footnote{Email address for correspondence: \url{michael.benzaquen@polytechnique.edu}}}
\affil[1]{Ladhyx, UMR CNRS 7646, Ecole polytechnique, 91128 Palaiseau Cedex, France}
\affil[2]{Chair of Econophysics \& Complex Systems, Ecole polytechnique, 91128 Palaiseau Cedex, France}
\affil[3]{Capital Fund Management, 23-25, Rue de l'Université 75007 Paris, France}
\date{\today \vspace{-1cm}}
\begin{document}

\maketitle

\begin{abstract}
We propose an actionable calibration procedure for general Quadratic Hawkes models of order book events (market orders, limit orders, cancellations). One of the main features of such models is to encode not only the influence of past events on future events but also, crucially, the influence of past {\it price changes} on such events. We show that the empirically calibrated quadratic kernel is well described by a diagonal contribution (that captures past realised volatility), plus a rank-one ``Zumbach'' contribution (that captures the effect of past trends). We find that the Zumbach kernel is a power-law of time, as are all other feedback kernels. As in many previous studies, the rate of truly exogenous events is found to be a small fraction of the total event rate. These two features suggest that the system is close to a critical point -- in the sense that stronger feedback kernels would lead to instabilities. 
\end{abstract}

\tableofcontents

\section{Introduction}

The accumulation of empirical clues over the past few years provides mounting evidence that most of market volatility is of endogenous nature~\cite{Cutler,Fair,joulin2008stock,bouchaud2010endogenous,fosset2019endogenous}. This obviously does not mean that significant news, such as the very recent Covid-19 crisis, do not impact financial markets, but rather that these only account for a small fraction of large price moves. Think for example of the S\&P500 flash crash of May 6th, 2010 \citep{kyle}, see also~\cite{zweig2010}, which has not been triggered by any outstanding piece of news. Furthermore, while one may argue that in some cases large drops are exogenously triggered, their amplification is often due to endogenous mechanisms \cite{bouchaud2018trades}. 

The behaviorally supported idea that agents tend to overreact, especially during crises, has driven the market modeling community to fall back on self-exciting processes, better known as Hawkes processes~\cite{hawkes1971spectra}. The latter have proven to be extremely efficient to tackle the intricate dynamics of the order flow and other self-excited effects in financial markets~\cite{toke2011introduction,bacry2015hawkes, morariu2018state, rambaldi2017role, bormetti2015modelling, wu2019queue, bacry2014hawkes, bacry2013modelling, hardiman2013critical, rambaldi2015modeling,huang2015simulating, alfonsi2016extension, alfonsi2016dynamic, achab2018analysis, calcagnile2018collective, filimonov2012quantifying,koyama2020statistical}. 
Nonetheless, linear Hawkes processes are unable to account for an empirical finding essential to our eyes to tackle endogenous instabilities: the Zumbach effect \citep{zumbach2010volatility,chicheportiche2014fine,blanc2017quadratic,el2020zumbach}. The latter states that past price trends increase future activity, regardless of their sign. Quadratic Hawkes processes (Q-Hawkes), inspired by quadratic ARCH processes~\citep{sentana1995quadratic,chicheportiche2014fine}, were recently introduced to circumvent this issue~\citep{blanc2017quadratic, dandapani2019quadratic}, and have proven key to understand fat-tails in the distribution of returns, as well as spread, volatility and liquidity dynamics \cite{fosset2019endogenous}.

In our recent paper~\cite{fosset2019endogenous} we indeed argued that price or spread jumps could be the result of endogenous feedback loops that trigger liquidity seizures, see also \cite{dall2019does}. In particular, we empirically showed that Zumbach-like effects exist in order book data, \textit{i.e.}~past trends and volatility tend to promote future activity, and in particular cancellations that diminish liquidity and fragilise the system, possibly leading to a liquidity crisis. Combining Q-Hawkes processes with a stylized order book model~\citep{daniels2003quantitative, smith2003statistical} revealed an interesting scenario with a second order phase transition between a stable regime for weak feedback and an unstable regime for strong feedback, in which liquidity crises arise with high probability. However, for such a scheme to be relevant for financial markets, the system must sit very close to the instability threshold (perhaps as the result of ``self-organised criticality''). As an alternative scenario, we also proposed a non-linear Hawkes process which exhibits liquidity crises as occasional ``activated'' events, separating locally stable periods of normal activity.

In the present paper, we calibrate on real market data a version of the generalized Q-Hawkes process proposed in our recent work~\cite{fosset2019endogenous}. We provide convincing evidence for the price/liquidity feedback mechanism described above and quantify its implications. In section~\ref{sec:general} we briefly recall  the ingredients of the model and present the non-parametric calibration procedure, inspired by the methods introduced by Bacry \textit{et  al.}~\citep{bacry2016first,bacry2016estimation, bacry2012non}. We apply such calibration to order book data on the EURO STOXX and BUND futures contracts. In section~\ref{sec:aggregate} we present an alternative method that needs fewer assumptions to compute the overall effect of past price moves on future liquidity flow. We introduce a low rank (Zumbach-like) approximation that allows us to denoise the feedback kernels and separate the effects of trend and volatility, and apply it to our futures contracts. In section~\ref{sec:spread}, we focus on the liquidity flow and analyse spread time series in relation with adequate trend and volatility signals. Results appear to favour the ``self-organized criticality'' scenario over the metastable, ``activated'' scenario discussed above and in~\cite{fosset2019endogenous}. In section~\ref{sec:conclusion} we conclude.

\section{Brute Force Calibration of a Q-Hawkes Process}\label{sec:general}

\subsection{Definition of the Model}

We present a simplified version of the Generalized Quadratic Hawkes process (GQ-Hawkes) introduced in~\citep{fosset2019endogenous}, where the influence of the size of the queues on event rates is neglected. 
Consider a 6-dimensional process $\boldsymbol{N}_t = \big( N_t^{\mathrm{C,b}} ,\ N_t^{\mathrm{LO,b}} , \ N_t^{\mathrm{MO,b}} ,\  N_t^{\mathrm{MO,a}} ,\  N_t^{\mathrm{LO,a}} ,\  N_t^{\mathrm{C,a}} \big)$ counting six types of order book events: limit orders (LO), cancellations (C), and market orders (MO), for both the bid (b) and ask (a) sides of the order book; we consider best quotes only. We further assume that the process $\boldsymbol{N}_t$ is coupled to the past price process $P_{t'< t}$ 
in the following way. Denoting $\boldsymbol{\lambda}_t$ the intensity of the the 6-dimensional process $\boldsymbol{N}_t$ we let:
\begin{equation}\label{eq:lambda}
\boldsymbol{\lambda}_t = \boldsymbol{\alpha}_0 + \int_0^t \boldsymbol{\phi} (t-s) \, d \boldsymbol{N}_s  + \int_0^t \boldsymbol{L} (t-s) \, d P_s 
			+  \int_0^t \int_0^t \boldsymbol{K} (t-s,t-u)  \, d P_s d P_u\,,
\end{equation}
with $\boldsymbol{\phi}$, $ \boldsymbol{L}  $ and $\boldsymbol{K} $ causal decaying kernels. One can always choose $\boldsymbol{K}(u,s)=\boldsymbol{K} (s,u)$ without loss of generality. Note that $\boldsymbol{\phi}$ is a 6$\times$6 matrix, whereas $ \boldsymbol{L}  $ and $\boldsymbol{K}$ are 6-dimensional vectors. 

The intensity $\boldsymbol{\lambda_t}$ is the sum of four different contributions, from left to right in the RHS of Eq.~\eqref{eq:lambda}, one has the base rate $\boldsymbol{\alpha}_0$, the standard linear Hawkes contribution, followed by the linear and the quadratic contributions of price fluctuations. As pointed out in \citep{fosset2019endogenous, blanc2017quadratic}, assuming that $P_t$ is a martingale makes analytical calculations, and numerical calibration, much more congenial. 
Finally, assuming as we shall do hereafter that a stationary state is reached allows us to replace the lower bound of the integrals in Eq.~\eqref{eq:lambda} by $-\infty$.

\subsection{A Non-Parametric Calibration Procedure}\label{section:calib}

Here we introduce a non-parametric scheme to calibrate Eq.~\eqref{eq:lambda} to real market data. Our method is an extension of the second moment method introduced by Bacry \emph{et al.} in~\citep{bacry2016first,bacry2016estimation}, see also~\cite{chicheportiche2014fine}.


\subsubsection{Covariances and Wiener-Hopf-like Equations}\label{section:set_eq}

Before deriving the equations that will be used for the calibration, we introduce the following averages and covariances:  
\begin{subeqnarray}\label{eq:defs}
   \Delta_k dt &:=& \mathbb{E} \left[ \left( d P_t\right)^k \right],\slabel{eq:deltapk}\\
  {\Lambda}^idt &:=& \mathbb{E} \left[ d {N}^i_t \right],\\
    \chi_{NN}^{ij}(t-s) \, dt ds  &:=&  \Cov \left(d N_t^i, d N_s^j \right) - \Lambda^j \delta_{ij} \delta(t-s) dt ds\,,\\
    \chi_{NP}^{i}(t-s) \, dt ds &:=&  \Cov \left(d N_t^i, d P_s \right)  , \\
   \chi_{NP^2}^{i}(t-s) \, dt ds &:=&  \Cov \left(d N_t^i,  d P_s^2 \right) ,\\
   \chi_{NPP}^{i}(t-s,t-x) \, dt ds dx &:=& \Cov \left(d N_t^i, d P_x d P_s \right) , \\
   \chi_{P^2P^2}(t-s) \, dt ds &:=& \Cov \left(d P_t^2, d P_s^2  \right) - \Delta_4 \delta(t-s) ds dt ,
\end{subeqnarray}
where we have assumed for simplicity that the jumps of $P$ and $\boldsymbol{N}$ are not simultaneous. Note that while price jumps can only occur if one order book event triggers them, the relative frequency of the latter is so much larger that this approximation is fully justified. 
 Combining Eq.~\eqref{eq:lambda} with Eqs~\eqref{eq:defs} yields the following  set of equations for the first and second moments of the processes. Introducing the notations $||f|| = \int_\mathbb{R} f(t) dt$ and $\boldsymbol{K}_{\mathrm{d}}(t):=\boldsymbol{K}(t,t)$ the diagonal part of $\boldsymbol{K}$, one obtains for $t,x > 0$ with $t \neq x$:
\begin{subeqnarray}\label{eq:gqhawkes_moments}
\Lambda^i & = & \alpha_0^i + \sum_k ||\phi^{ik}|| \Lambda^k + ||K_{\mathrm{d}}^i|| \Delta_2, \slabel{eq:mean_intensity}
\\
 \chi_{NN}^{ij}(t) & = & \Lambda^j \phi^{ij}(t) +  \int_{\mathbb{R}^+} \left[\sum_k \phi^{ik}(s) \chi_{NN}^{kj}(t-s) +  L^i (s) \chi_{NP}^{j}(s-t) +  K_{\mathrm{d}}^i(s) \chi_{NP^2}^{j}(s-t) \right] ds,  \slabel{eq:cov_event_event}\nonumber
\\ 
&&  + \int_{[t,+ \infty[^2}  K^i(s,u) \chi_{NPP}^{j}(s-t,u-t) 1_{\{ s \neq u \}} du ds,
\\
 \chi_{NP}^{i}(t) & = &  \int_{\mathbb{R}^+} \sum_k \phi^{ik}(s) \chi_{NP}^{k}(t-s) ds  + L^i (t) \Delta_2  + K_{\mathrm{d}}^i(t) \Delta_3, \slabel{eq:cov_event_P}
\\
 \chi_{NP^2}^{i}(t) & = & \int_{\mathbb{R}^+} \sum_k \phi^{ik}(s) \chi_{NP^2}^{k}(t-s) ds + L^i (t) \Delta_3 + K_{\mathrm{d}}^i(t)  \Delta_4 + \int_{\mathbb{R}^+} \chi_{P^2P^2}(t-s) K_{\mathrm{d}}^i(s) ds , \slabel{eq:cov_event_sqP}
\\
 \chi_{NPP}^{i}(t,x) & = & \int_{\mathbb{R}^+} \sum_k \phi^{ik}(s) \chi_{NPP}^{k}(t-s,x-s) ds + 2 \Delta_2^2  K^i(t,x). \slabel{eq:cov_event_PP}
\end{subeqnarray}
Provided the number of events generated by price fluctuations is small compared to that generated by the linear Hawkes contribution, \textit{i.e.} $\sum_{i,k} ||\phi^{ik}|| \Lambda^k \gg \sum_i ||K_{\mathrm{d}}^i|| \Delta_2$,  Eq.~\eqref{eq:cov_event_event} conveniently simplifies to:
\begin{equation} \label{eq:cov_hawkes}
    \chi_{NN}^{ij}(t) = \Lambda^j \phi^{ij}(t) + \sum_k \int_{\mathbb{R}^+} \phi^{ik}(s) \chi_{NN}^{kj}(t-s) ds.
\end{equation}
 This approximation is relatively well supported by real data for short enough times (see below). It is essential at this stage as it allows us to decouple the estimation of the Hawkes kernel from that of $ \boldsymbol{L}  $ and $\boldsymbol{K} $:  one can first estimate $\boldsymbol{\phi}$ from Eq.~\eqref{eq:cov_hawkes} and then compute  $\boldsymbol{L}$ and $\boldsymbol{K}$ from Eqs.~\eqref{eq:cov_event_P}, \eqref{eq:cov_event_sqP} and \eqref{eq:cov_event_PP}. The base rate is finally obtained from Eq.~\eqref{eq:mean_intensity}. Note that while in principle an exact calibration of Eqs.~\eqref{eq:gqhawkes_moments} is possible, it does not perform well on real data -- but see section \ref{sec:aggregate} below.

\subsubsection{Micro-Price, Discretisation and Calibration Recipe}\label{section:procedure}

In section~\ref{sec:general} we stressed that the point process $P_t$ needs to be a martingale for Eqs.~\eqref{eq:gqhawkes_moments} to be valid. Yet, it is well established that the mid-price in financial markets displays substantial mean-reversion at short timescales. To circumvent this issue we consider the volume weighted mid-price, sometimes called the \emph{micro-price}, $P_t^\text{micro}$, known to be closer to a martingale at high frequency \cite{stoikov2018micro, gould2016queue}.\footnote{More refined definitions of the micro-price, even closer to a martingale, are discussed in \cite{stoikov2018micro}.} 
It is defined as:
\begin{equation}\label{eq:def_vw_price}
    P_t^\text{micro} = \frac{v^{\mathrm{a}}_t b_t + v^{\mathrm{b}}_t a_t}{v^{\mathrm{a}}_t + v^{\mathrm{b}}_t},
\end{equation}
where  $v^{\mathrm{b}}_t$, $v^{\mathrm{a}}_t$ denote the available volume at the best bid $b_t$ and ask $a_t$ respectively. To enforce further the martingale property we use the so-called \emph{surprise} price, that we shall henceforth denote by $P_t$, and which consists in subtracting to the price its (linear) statistical predictability. Mathematically speaking, this reads:
\begin{equation}
    d P_t = d P_t^\text{micro} - \int_{- \infty}^{t^-} \rho_P(t-s) d P_s^\text{micro},\label{eq:surprise}
\end{equation}
where $\rho_P(t-s):= \Cor \left(dP_t^\text{micro}, dP_s^\text{micro}\right)$ denotes the price auto-correlation function.

We also note that the intensity of order book events exhibit an intraday U-shape, very much like the well known U-shaped volatility pattern. Computing the total intensity of events $\Lambda_{\mathrm{tot}} = \sum_i \Lambda_i$ over 5-min bins and averaging over trading days, a U-shape is clearly visible. To avoid spurious effects related to these intraday seasonalities, we rescale time flow  by this average pattern to enforce a constant rate of events in the new time variable.
 
In order to estimate the kernels from real order book data, one must  choose a  time grid $t^H_n $ with weights $w^H_n$ for kernel $\boldsymbol{\phi}$, such that $|| \boldsymbol{\phi} || \approx \sum_n  \boldsymbol{\phi} (t^H_n) w^H_n$. We decide to use quadrature points~\cite{bacry2016estimation} to ensure a good approximation of the integrals with a minimal number of points. Further, given that we expect power-law kernels, see e.g.~\cite{hardiman2013critical,bacry2016estimation, blanc2017quadratic}, we choose a linear scale at short times that switches to logarithmic at longer times. Finally, given that typical timescales are usually quite different (see below), we choose a different time grid $t_n$, $w_n$ for the kernels $\boldsymbol{L}$ and $\boldsymbol{K}$. See Appendix~\ref{appendix:estimation} for more details. 

Finally, the empirical covariances are usually very noisy, so we choose to smooth them using a convenient fitting function in order to obtain better behaved kernels. Concerning the volatility covariance $\chi_{P^2P^2}$, it is found to behaves like a power law at large times so the chosen fitting function is\footnote{For the EUROSTOXX, the fitting parameters are found to be $A = 1.7 \times 10^{-4}\,\$^4 s^{-2}$, $B = 81$\,s and $C = 0.71$.} $A (1 + t / B)^{-C}$ We also fit the logarithm of $\boldsymbol{\chi}_{NP}(t)$, $\boldsymbol{\chi}_{NP^2}(t)$ by a polynomial in $\log t$, and smooth the off-diagonal kernel, see \ref{section:zumbach_contribution} for details. Plots of the ``raw'' kernels obtained without smoothing fits are provided in Appendix.~\ref{appendix:plots}. Apart from being more noisy, as expected, these raw kernels are very similar to the smoothed ones.\pagebreak 

The calibration recipe then amounts to the following steps.
\begin{itemize}\itemsep-0.3em 
    \item Compute the surprise price from the micro-price using Eqs~\eqref{eq:def_vw_price} and \eqref{eq:surprise}.
    \item Rescale  time by the typical daily pattern of $\Lambda_{\mathrm{tot}} = \sum_i \Lambda_i$.
    \item Estimate $\Delta_k$, $\boldsymbol{\Lambda}$ and the covariances $\chi_{P^2P^2}$, $\boldsymbol{\chi}_{NN}$, $\boldsymbol{\chi}_{NP}$, $\boldsymbol{\chi}_{NP^2}$ and $\boldsymbol{\chi}_{NPP}$ from the data using Eqs~\eqref{eq:defs},
    \item Use adequate fitting functions to smooth the empirical covariances (optional),
    \item Discretise and solve Eq.~\eqref{eq:cov_hawkes} to obtain the Hawkes kernel $\boldsymbol{\phi}$,
    \item Discretise and solve Eqs.~\eqref{eq:cov_event_P},~\eqref{eq:cov_event_sqP}, and~\eqref{eq:cov_event_PP} to obtain the kernels $\boldsymbol{L}$ and $\boldsymbol{K}$,
    \item Discretise and solve Eq.~\eqref{eq:mean_intensity} to obtain the base rate $\boldsymbol{\alpha}_0$. 
\end{itemize}{}
Further details on how to solve these equations in practice are provided in Appendix~\ref{appendix:estimation}.

\subsection{Empirical Results}

\begin{figure}[t!]
  \centering
  \includegraphics[width=0.45\columnwidth]{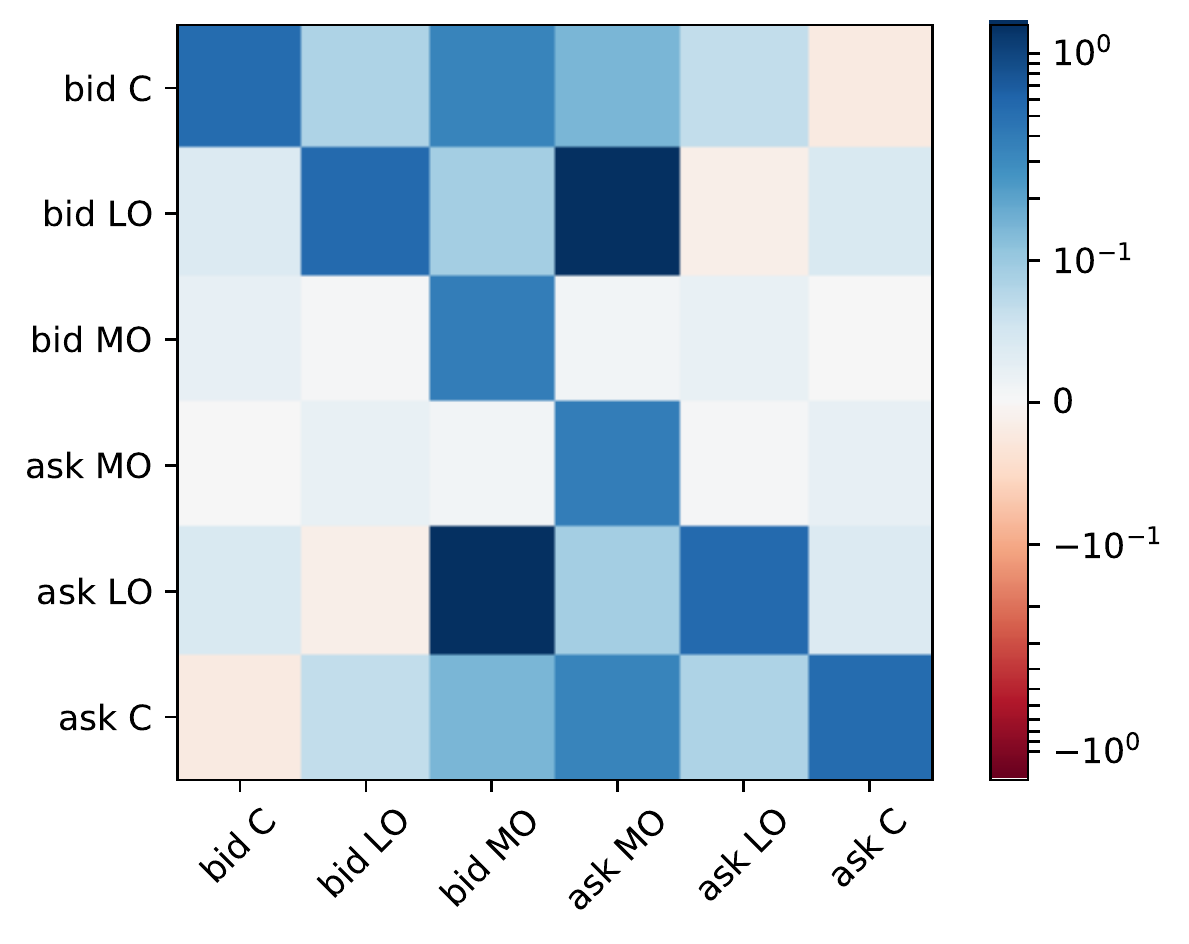}
  \caption{Norms of the Hawkes kernel $||\phi^{ij}||$ for the EURO STOXX futures contract between $2016/09/12$ and $2020/02/07$, calibrated using Eq. \eqref{eq:cov_hawkes}. }
  \label{fig:calibEurostock_hawkes}
\end{figure}

We now apply the calibration procedure presented above to the  EURO STOXX futures  contract in the period $2016/09/12$ to $2020/02/07$. For this contract, the average time between two order book events is $\tau_{\text{e}} \approx 0.03s$, two orders of magnitude below the average time between two price changes $\tau_P \approx 7s$, indicating that the range of the kernels $\boldsymbol{L}$ and $\boldsymbol{K}$ is likely to be greater than that of $\boldsymbol{\phi}$, and allowing one to choose discretisation time grids accordingly. We also apply the procedure to the BUND futures contract but do not show all the (redundant) results for the sake of readability; summarising results are displayed in Fig.~\ref{fig:quadratic_contribution_hist} and Tables~\ref{tableVolume}, \ref{tableQuadraticContributions} and \ref{tableVolatility}. 

As specified in section~\ref{section:procedure}, we start with the calibration of the Hawkes kernel $\boldsymbol \phi$. The results are displayed in Fig.~\ref{fig:calibEurostock_hawkes} for the norms of the kernels, and in Fig.~\ref{fig:calibEurostock_hawkes_kernels} in the Appendix for the full time-dependence.  The temporal decay of the kernels appears to be power law with exponent $\approx -1.5$, consistent with previous reports~\cite{hardiman2013critical,bacry2016estimation, blanc2017quadratic}.

\begin{figure}[t!]
  \centering
  \includegraphics[width=0.9\columnwidth]{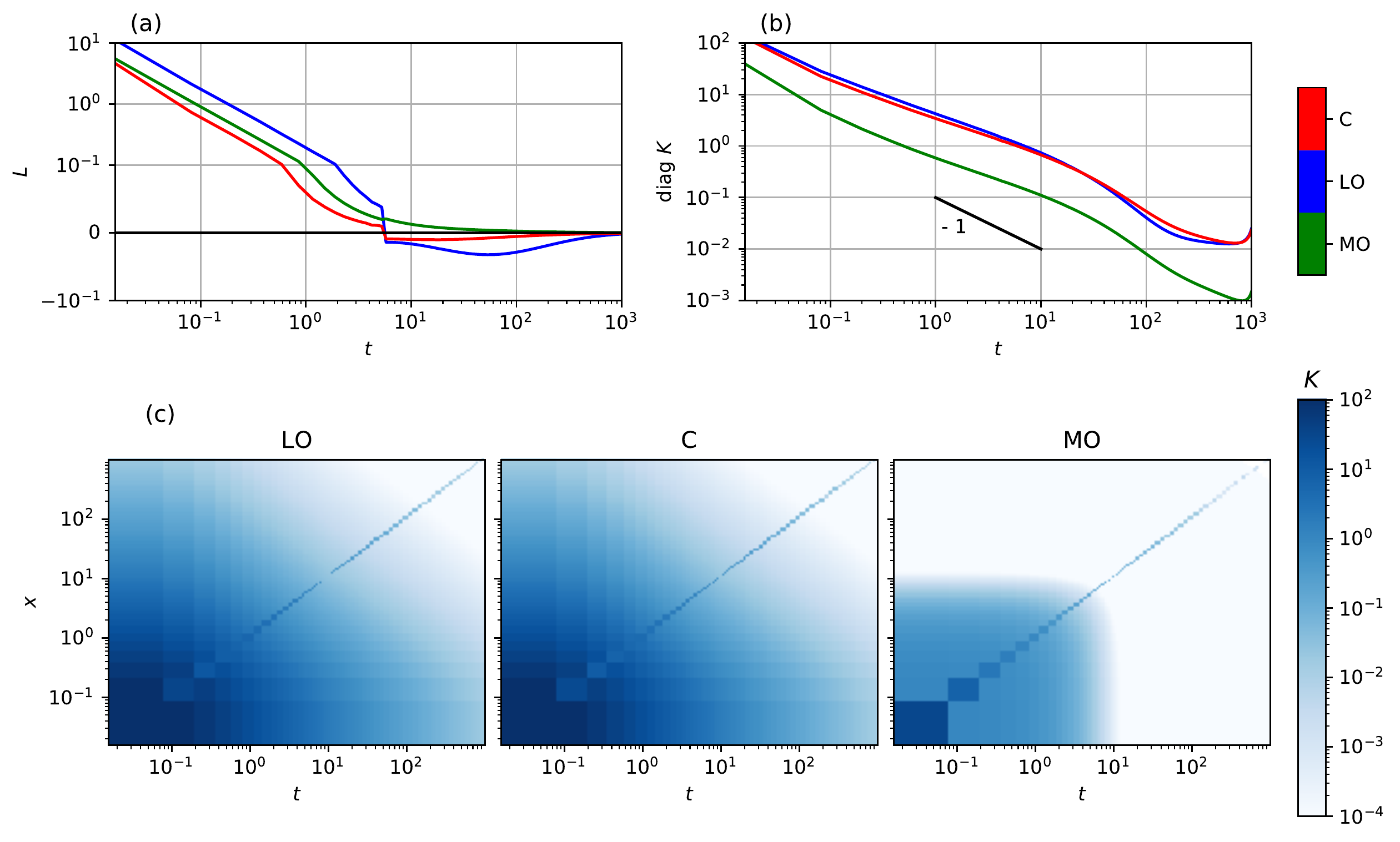}
  \caption{Kernels resulting from the non-parametric calibration on the EURO STOXX futures contract between $2016/09/12$ and $2020/02/07$. (a) Linear kernels $\boldsymbol L$. Note that the sign is such that an up (resp. down) trend increases all the event rates at the bid (resp. ask) at short times. (b) Diagonal of quadratic kernels $\boldsymbol{K}_{\mathrm{d}}$. (c) Full quadratic kernels $\boldsymbol{K}(t,x)$. }
  \label{fig:calibEurostock_kernels}
\end{figure}

The calibration leads to a stable Hawkes process with spectral radius of $||\mathbf{\phi}||$ (computed over 1000s) found to be $\approx 0.75$ for the EURO STOXX contract and $\approx 0.74$ for the BUND \citep{bacry2015hawkes,toke2011introduction}.  The results show that the expected bid-ask symmetry holds with a high level of accuracy (see~\cite{fosset2019endogenous}), such that one can average the kernels accordingly to improve the statistics without loss of information.

Plugging the obtained Hawkes kernels into Eqs.~\eqref{eq:cov_event_P},~\eqref{eq:cov_event_sqP} and~\eqref{eq:cov_event_PP} 
allows us to calibrate the kernels $\boldsymbol{L}$ and $\boldsymbol{K}$, see Fig.~\ref{fig:calibEurostock_kernels}. Again the expected bid-ask symmetry properties hold rather well: while the linear kernel $\boldsymbol L$ is anti-symmetric (the effect of the positive trend on the bid is the same as that of a negative trend on the ask), the quadratic kernel $\boldsymbol K$ is bid-ask symmetric.  We will therefore not distinguish further bid and ask events in the following. 

Figure~\ref{fig:calibEurostock_kernels}(c) shows that the quadratic contribution cannot be reduced to the diagonal part $\boldsymbol{K}_{\mathrm{d}}$ only. Indeed, the off-diagonal contribution of the kernel is non-zero and rather long-ranged. The decay of the diagonal contribution is a power law with exponent $\approx -1$. Such a decay is very slow and means that $||\boldsymbol{K}_{\mathrm{d}}||$ is logarithmically sensitive to long timescales, for which we do not have much information since we only use data belonging to the same trading day to avoid the thorny discussion of overnight effects and how to treat them. 

Finally, while the Hawkes and price feedback effects are difficult to compare as they do not operate on the same timescales, one can argue that the approximation presented at the end of Sec.~\ref{section:set_eq} is  well supported by data: considering a cut-off of 1000 seconds to compute the norms, one finds: $ \sum_i ||K_{\mathrm{d}}^i|| \Delta_2 / \sum_{i,k} ||\phi^{ik}|| \Lambda^k \approx 0.06$. 
Another useful piece of information is the global effect of the quadratic term on order book events, measured by $\sum_i ||K_{\mathrm{d}}^i|| \Delta_2$, which must be compared to the total activity $\sum_i \Lambda^i$. The ratio of these two quantities is found to be $5 \%$ for the EURO STOXX and $7 \%$ for the BUND (see Table \ref{tableVolatility} for more details). Although not dominant, this feedback is clearly not negligible. Together with the standard Hawkes contribution, this means that the exogenous contribution $\alpha$ to the total activity is only $19 \%$ of the total for the EURO STOXX ($17 \%$ for the BUND). Note that this fraction is expected to decreases further as the upper cut-off of the slowly decaying kernels is extended beyond 1000 seconds (see e.g. \cite{hardiman2013critical}). 

\section{A Simplified Framework}\label{sec:aggregate}

Here we present a framework which improves the above calibration in a threefold manner. As we shall see, (i) it allows to circumvent the approximation given in Eq.~\eqref{eq:cov_hawkes} which, we recall, is not perfectly satisfied by real data, (ii) it helps cleaning further the noisy off-diagonal contribution of the quadratic kernel, and (iii) it gives a more relevant measure of the global effect of price fluctuations on event rates with no longer having to consider, nor calibrate, the Hawkes contribution. 

\subsection{Effective Kernels}

\begin{figure}[t!]
  \centering
  \includegraphics[width=0.9\columnwidth]{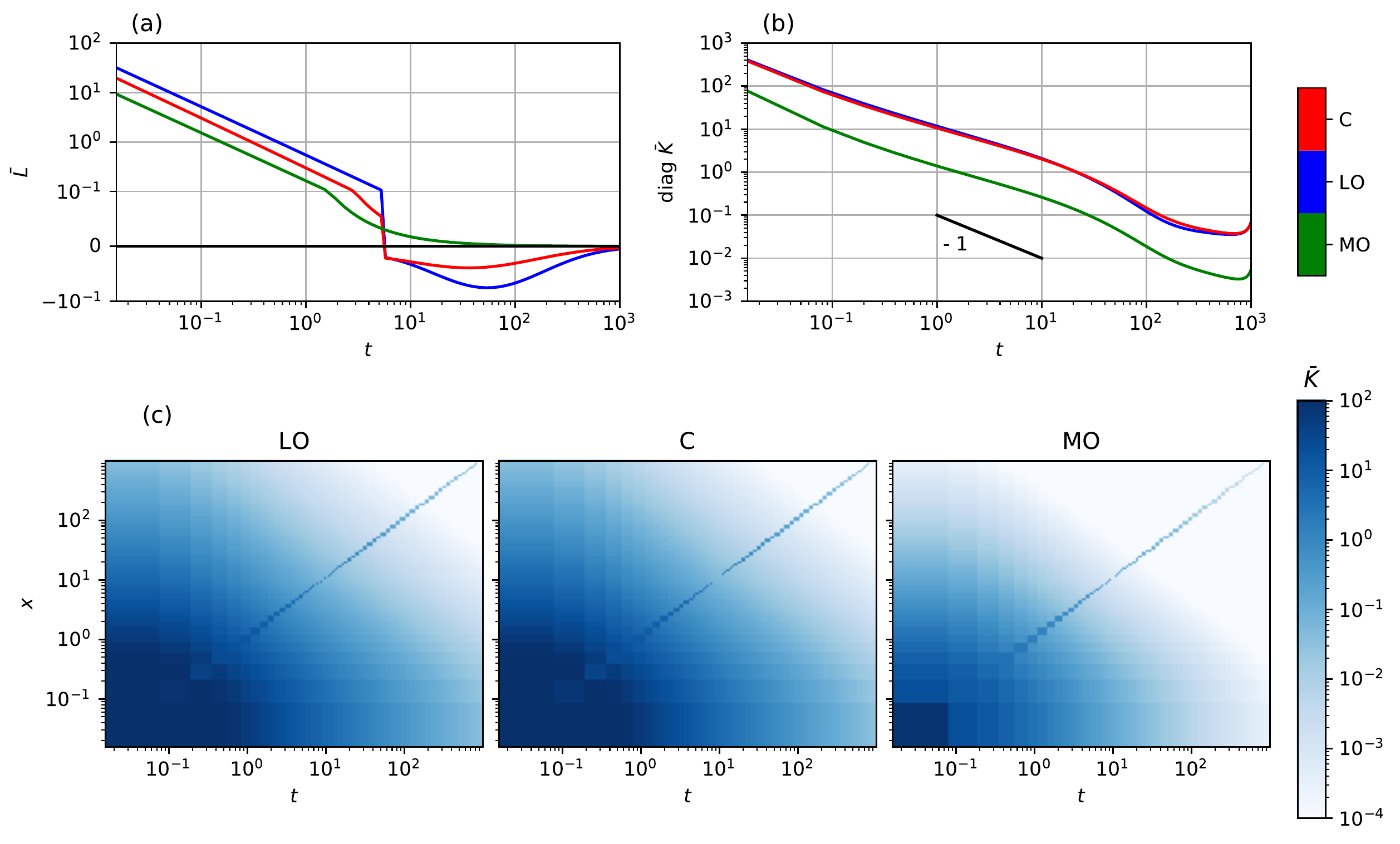}
  \caption{Effective kernels resulting from the simplified calibration on the EURO STOXX futures contract between $2016/09/12$ and $2020/02/07$. (a) Linear kernels $\boldsymbol{ \bar L}$. Note that the sign is such that an up (resp. down) trend increases all the event rates at the bid (resp. ask) at short times. (b) Diagonal of quadratic kernels $\boldsymbol{\bar K}_{\mathrm{d}}$.  (c) Full quadratic kernels $\boldsymbol{\bar K}(t,x).$}
  \label{fig:Eurostock_aggkernels}
\end{figure}

Using the resolvent method, see~\cite{bacry2016estimation, jaisson2015limit}, one can rewrite Eq.~\eqref{eq:lambda} as:
\begin{equation}\label{eq:lambda_agg_h}
\boldsymbol{\lambda}_t = \left(\mathbf{I} - || \boldsymbol{\phi} || \right)^{-1} \boldsymbol{\alpha}_0 + \int_{- \infty}^t \boldsymbol{\mathcal{R}} (t-s) \, d \boldsymbol{M}_s  + \int_{- \infty}^t \boldsymbol{\bar{L}} (t-s) \, d P_s 
			+  \int_{- \infty}^t \int_{- \infty}^t \boldsymbol{\bar{K}} (t-s,t-u)  \, d P_s d P_u ,
\end{equation}
with $ \boldsymbol{M}$ a martingale satisfying $d \boldsymbol{M}_t = d \boldsymbol{N}_t - \boldsymbol{\lambda}_t d t $, $ \boldsymbol{\mathcal{R}} = \sum_{n \geq 1} \boldsymbol{\phi}^{*n}$ the resolvent, $\boldsymbol{\bar{L}} = \boldsymbol{L} + \boldsymbol{\mathcal{R}} * \boldsymbol{L} $ and $\boldsymbol{\bar{K}} (t,s) = \boldsymbol{K} (t,s) + \int_0^{+ \infty} \boldsymbol{\mathcal{R}}(u) \boldsymbol{K} (t-u,s-u) d u $. The kernels $\boldsymbol{\bar{L}}$ and $\boldsymbol{\bar{K}}$ account for the overall feedback effect of $P_t$, including all subsequent Hawkes self-excited events that are induced by price fluctuations. The remarkable property of such kernels is that they solve a much simpler set of equations:
\begin{subeqnarray}\label{eq:simplerset}
 \chi_{NP}^{i}(t) & = & \bar{L}^i (t) \Delta_2  + \bar{K}_{\mathrm{d}}^i(t) \Delta_3 \slabel{eq:cov_event_P_h}
\\
 \chi_{NP^2}^{i}(t) & = & \bar{L}^i (t) \Delta_3 +  \bar{K}_{\mathrm{d}}^i(t)  \Delta_4 + \int_{\mathbb{R}} \chi_{P^2P^2}(t-s) \bar{K}_{\mathrm{d}}^i(s) ds \slabel{eq:cov_event_sqP_h}
\\
 \chi_{NPP}^{i}(t,x) & = & 2 \bar{K}^i(x,t) \Delta_2^2 \slabel{eq:cov_event_PP_h},
\end{subeqnarray}
where we have again enforced that $\boldsymbol{\bar{K}}$ is symmetric. The results obtained from the inversion of Eqs~\eqref{eq:simplerset} for the  EURO STOXX futures contract  are displayed in Fig.~\ref{fig:Eurostock_aggkernels}.  These lead to similar, though slightly cleaner, conclusions to Fig.~\ref{fig:calibEurostock_kernels}. In particular, the values of $\sum_i ||\bar{K}_{\mathrm{d}}^i|| \Delta_2$ are compatible with those obtained above (taking into account the $1 - || \boldsymbol{\phi} ||$ factor, see Table \ref{tableVolatility}). 

\subsection{The Zumbach Factorisation }\label{section:zumbach_contribution}

Here we further dissect the results of the calibration presented in the previous section, with the objective in particular of separating the contributions of trend and of volatility to the quadratic feedback. A meaningful approximation for the quadratic kernel $\boldsymbol{\bar K}$ was introduced in~\cite{blanc2017quadratic}, as the sum of a purely diagonal matrix and a rank-one contribution:\footnote{The the slight abuse of notation here since the diagonal part of $\bar K(s)$ is in fact $\bar{K}_{\mathrm{d}} \psi(s) + \bar{K}_1 Z^2(s)$.}
\begin{eqnarray}\label{eq:zapprox}
\bar K^i(t-s,t-u) := \bar{K}^i_{\mathrm{d}} \psi^i(t-s) \mathds{1}_{\{s = u\}} + \bar{K}^i_1 Z^i(t-s)Z^i(t-u) \, .
\end{eqnarray}
The first term on the right hand side of Eq.~\eqref{eq:zapprox} reflects feedback of past volatility on current order book events. Its contribution in Eq.~\eqref{eq:lambda_agg_h} can indeed by written as: 
\begin{eqnarray}\label{eq:def_vol}
    \left[\sigma^i(t) \right]^2 := \int_0^t \psi^i(t-s)  \, (d P_s)^2  ,
\end{eqnarray}
    The second term  is in turn a reflection of the effect of past trends, as measured in Eq.~\eqref{eq:lambda_agg_h} by $[\mu^i(t)]^2$, where: 
\begin{eqnarray}\label{eq:def_trend}
    \mu^i(t):=\int_0^t Z^i(t-s)  \, d P_s . 
    \end{eqnarray}
    This last term is reminiscent of the so-called Zumbach effect: past trends, regardless of their sign, lead to an increase in future activity. An altenative interpretation is that $[\mu^i(t)]^2$ is a local measure of a low-frequency volatility, to be contrasted with $[\sigma^i(t)]^2$ which is a local measure of high-frequency volatility. Note that the kernels $\psi$ and $Z$ are normalised:
    \begin{equation} \label{eq_normalisation}
    \int \psi^i(s) ds = \int Z^i(s)^2 ds = 1,
    \end{equation}
    such that the overall strength of the volatility contribution is $\bar{\boldsymbol{K}}_{\mathrm{d}} $ while that of the trend contribution is $\bar {\boldsymbol{K}}_1$.
    
\begin{figure}[t!]
  \centering
  \includegraphics[width=0.9\columnwidth]{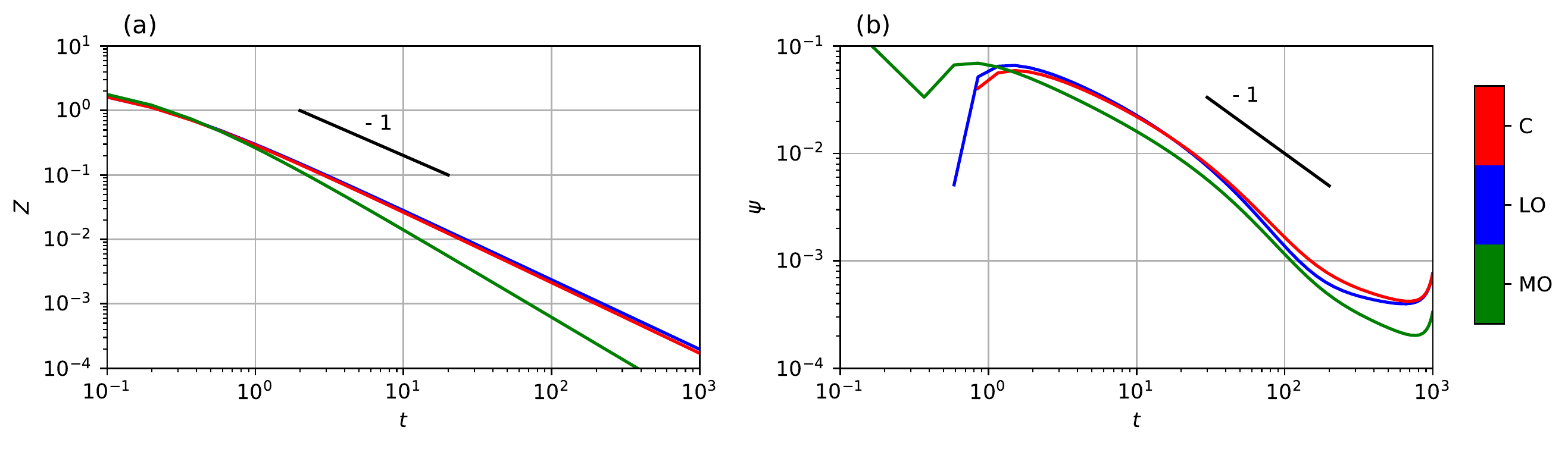}
  \caption{Zumbach approximation of the effective kernel $\bar{\boldsymbol{K}}$ on the EURO STOXX futures contract between $2016/09/12$ and $2020/02/07$. (a) Zumbach kernel ${\boldsymbol{Z}}$, (b) Volatility kernel ${\boldsymbol{\psi}}$.  Both kernels are normalised such that $||{\boldsymbol{\psi}}|| = || {\boldsymbol{Z}}^2 || =1$, with a cut-off in the time integrals at 1000 secs.}
  \label{fig:zumbach_vol_kernels}
\end{figure}
    
    While in practice such an approximation is of course not perfect, one can check that including higher rank contributions is unessential as the latter do not carry much additional signal. The rank-one kernel is obtained by  minimizing  {$\iint \left(\bar K^i(s,u) -  \bar{K}^i_1 Z^i(s)Z^i(u) \right)^2 \mathds{1}_{\{u \neq s \}}ds du$}, which consists in finding the first eigenvector of a well chosen linear map, see \cite{srebro2003weighted} for more details. The $\psi$ contribution is then obtained by taking the diagonal of $\bar K^i$ and subtracting $\bar{K}^i_1 Z^i(t)^2$. Figure~\ref{fig:zumbach_vol_kernels}
    displays the kernels $\boldsymbol{\phi}$ and $\boldsymbol{Z}$ as function of time for the EUROSTOXX  futures contract. As one can see, while the volatility kernel decays roughly as $1/t$, although some curvature can be observed. The Zumbach counterpart decays as $1/t$, regardless of event types (by that justifying the choice made in Fosset \textit{et al.}~\cite{fosset2019endogenous}, where the same functional form for all event types was assumed).

\section{Liquidity Dynamics \& Crises}\label{sec:spread}
 
\subsection{Quadratic Feedback on Liquidity}
   
\begin{figure}[t!]
  \centering
  \includegraphics[width=\columnwidth]{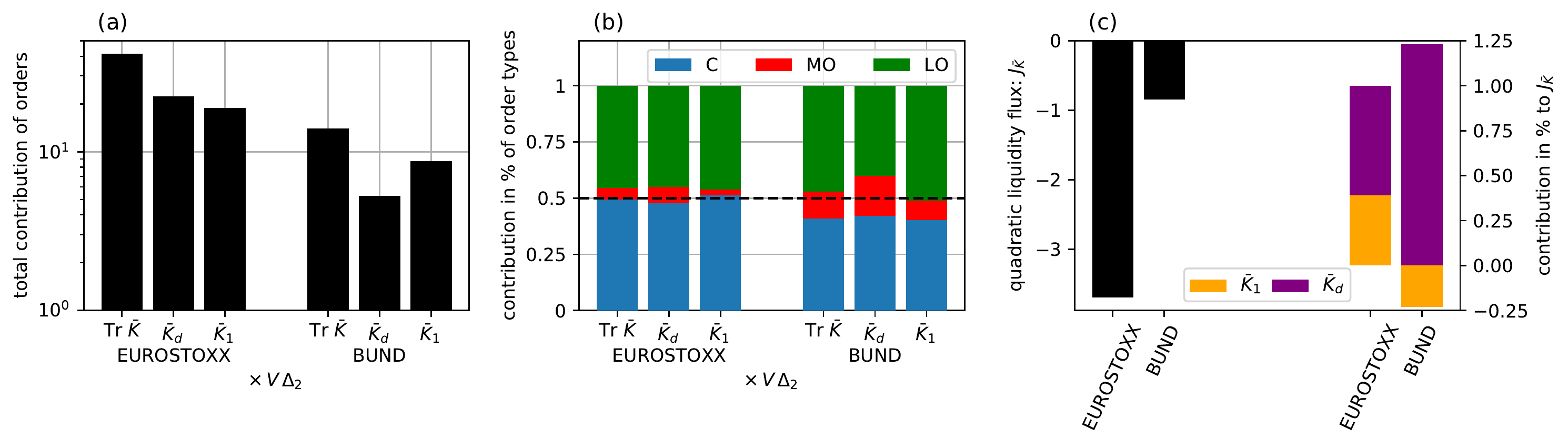}
  \caption{Average quadratic contribution on the EURO STOXX and BUND futures contracts between $2016/09/12$ and $2020/02/07$. (a) $\Delta_2 \sum_i V^i ||\bar K^i||$ and its decomposition into $\Delta_2 \sum_i V^i \bar{K}_{\mathrm{d}}^i$ and $\Delta_2 \sum_i V^i \bar{K}_1^i$. (b) Contributions of each order type to the latter quantities. (c) Overall contribution of the quadratic effect to the liquidity flow $J_{ \boldsymbol {\bar K}}$ (in shares per second), and relative contribution of the volatility and Zumbach terms.}
  \label{fig:quadratic_contribution_hist}
\end{figure}

So far we have focused on the impact of past price moves one event rates. Here we wish to go on step further and estimate the effect of past price changes on liquidity, \emph{i.e.} volume weighted events. For this one needs to consider order volumes.  The average volumes  are given in Tab.~\ref{tableVolume} for the different types of orders.
\begin{table}[h!]
\begin{centering}
\begin{tabular}{|c|c|c|c|c|c|c|}
\hline 
 &  \small $V^{\mathrm{C,b}}$ &  \small $V^{\mathrm{LO,b}}$ &  \small $V^{\mathrm{MO,b}}$ &  \small $V^{\mathrm{MO,a}}$ & \small $V^{\mathrm{LO,a}}$ & \small $V^{\mathrm{C,a}}$   \\ \hline
\small EUROSTOXX  & \small  $10.1$  & \small $9.2$ & \small   $7.2 $  & \small $8.2$ & \small $9.2$   & \small   $10.0$  \\ \hline
\small BUND  & \small    $4.5$ & \small $4.8 $ & \small  $4.4$   & \small  $ 4.2 $ & \small $4.8$ & \small   $4.5$  \\ \hline
\end{tabular}
\caption{Average order volumes (in shares).}
\label{tableVolume}
\end{centering}
\end{table}
Assuming bid/ask symmetry (consistent with the empirical results), Fig.~\ref{fig:quadratic_contribution_hist} displays the amount of shares per second that can be attributed to the quadratic effect (both volatility and Zumbach) for each event type, namely $\bar{K}_{\mathrm{d}}^i V^i \Delta_2$ and $\bar{K}_1^i V^i \Delta_2$ where $\bar{K}_{\mathrm{d}}^i,\bar{K}_1^i$ are obtained as explained in the previous section, $V^i$ are given in Tab.~\ref{tableVolume}, and $\Delta_2$ is defined in Eq.~\eqref{eq:deltapk}.\footnote{The normalisation of all kernels is computed with a time cut-of at 1000 seconds.}

Introducing the overall average quadratic liquidity flux as: 
\begin{eqnarray}
  J_{ \boldsymbol {\bar K}} := \left(||\bar K^{\mathrm{LO}}|| V^{\mathrm{LO}} -  ||\bar K^{\mathrm{C}}|| V^{\mathrm{C}} - ||\bar K^{\mathrm{MO}}|| V^{\mathrm{MO}}\right) \Delta_2,  
\end{eqnarray}
one consistently finds  that the quadratic (price) feedback has an overall negative effect on liquidity  $J_{ \boldsymbol {\bar K}}< 0$, most of it associated to volatility, see Fig.~\ref{fig:quadratic_contribution_hist}(c).
\footnote{Note that the linear terms give no net contribution, i.e.  $V^{\mathrm{LO}} ||\bar{L}^{\mathrm{LO}}|| - V^{\mathrm{C}} ||\bar{L}^{\mathrm{C}}|| - V^{\mathrm{MO}} ||\bar{L}^{\mathrm{MO}}|| \approx 0$, which explains why we focus on the quadratic term). In other words, the trend has almost no linear effect on the liquidity flux at large time scales.}
In other terms, the quadratic feedback tends to decrease liquidity on average. Figure~\ref{fig:quadratic_contribution_hist}(b) shows that both the volatility and Zumbach terms have an average negative impact on liquidity (i.e. the green bars represent less than $50 \%$ of the total contribution).
The Zumbach term is responsible for non-trivial long-range liquidity anomalies. In particular, Blanc \emph{et al.}~\cite{blanc2017quadratic} showed that the price process resulting from a quadratic Hawkes process follows is diffusive with fat tailed stochastic diffusivity at large times, which can be attributed to the Zumbach effect, rather than its volatility counterpart (see also the discussion in \cite{dandapani2019quadratic}). In any case, we believe that the quadratic feedback of price trends on order book events is a crucial ingredient to understand liquidity crises. In the next section we provide a direct test of this hypothesis.

\subsection{Spread Dynamics and Liquidity Crises}

\begin{figure}[t!]
  \centering
  \includegraphics[width=0.98\columnwidth]{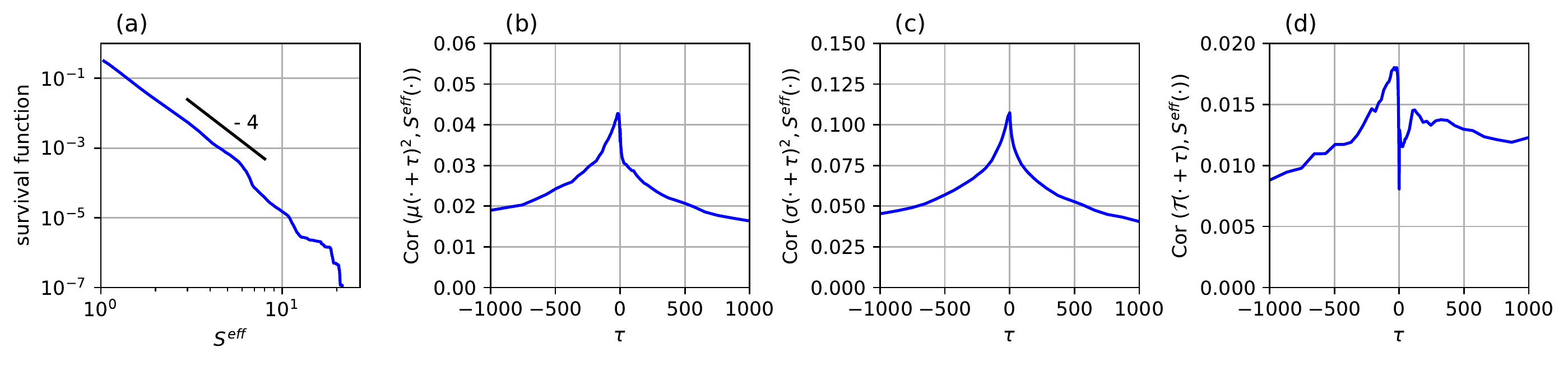}
  \caption{(a) Survival function of the effective spread, showing that $\mathbb{P}(S^\mathrm{eff} > S) \sim S^{-4}$  (b) Correlations between effective spread and past square trends (for $\tau < 0$) and future square trends (for $\tau > 0$).(c) Correlations between effective spread and square volatility. (d) Correlations between effective spread and $\mathcal{T}$: the ratio square trend over square volatility. EURO STOXX futures contracts between 2016/09/12 and 2020/02/07. 
}
  \label{fig:corr}
\end{figure}

With the aim of making contact with our previous work~\cite{fosset2019endogenous}, we now focus on the analysis of spread dynamics. Since the EUROSTOXX futures is a large tick contract (the spread is equal to one over $99\%$ of the time and seldom higher than two), we characterize the dynamics of liquidity using an \emph{effective spread} $S^\mathrm{eff}_t$ which is defined as follows.  Calling $v^{\mathrm{a}}_t(x)$ (\textit{resp} $v^{\mathrm{b}}_t(x)$) the ask (\textit{resp} bid) volume at price level $x$,  we construct cumulative volumes as $Q^{\mathrm{a}}_t(x)  = \sum_{n \leq x} v^{\mathrm{a}}_t(n) $ and $Q^{\mathrm{b}}_t(x)  = \sum_{n \geq x} v^{\mathrm{b}}_t(n) $. We then choose the average volume at best $V_\mathrm{best}$ as a reference volume, and define:\footnote{Changing the reference volume to $2V_\mathrm{best}$ or $V_\mathrm{best}/2$ does not change the qualitative conclusions below.}
\begin{equation}
S^\mathrm{eff}_t := \left( Q^{\mathrm{a}}_t \right)^{-1} \left(V_\mathrm{best} \right) - \left( Q^{\mathrm{b}}_t \right)^{-1} \left(V_\mathrm{best} \right),
\end{equation}
where $\big( Q^{\mathrm{a/b}}_t \big)^{-1}$ denotes the inverse function of $Q^{\mathrm{a/b}}_t$. The effective spread is a natural proxy for liquidity in the close vicinity of the midprice: when the liquidity is close to its average, the effective spread coincides with the regular spread; but when liquidity is low, it can be much larger as aggregating the volume of several queues is needed to recover the reference  volume $V_\mathrm{best}$. Figure~\ref{fig:corr}(a) displays the survival function of the effective spreads, revealing that $\mathbb P(S^{\text{eff}})\sim (S^{\text{eff}})^{-5}$. This power-law tail is interesting for the following reason: the effective spread can be seen as a proxy for the size of latent price jumps, i.e. the jumps that are likely to happen if an aggressive market order hits the market. Hence, one expects the distribution of effective spread is not far from the distribution of price returns $r$, which is well known to decay as $\mathbb P(r)\sim r^{-4}$.

Let us now study the relation between effective spread, square volatility $\sigma^2$ and square trend $\mu^2$, as  defined in Eqs.~\eqref{eq:def_vol} and \eqref{eq:def_trend}. Figures~\ref{fig:corr}(b), (c) and (d) display the correlation functions $C_\mu(\tau):=\Cor \left[\mu(t+\tau)^2, S^\mathrm{eff}(t) \right]$,  $C_\sigma(\tau):=\Cor \left[\sigma(t+\tau)^2, S^\mathrm{eff}(t) \right]$ and  $C_\mathcal{T}(\tau):=\Cor \left[\mathcal{T}(t+\tau), S^\mathrm{eff}(t) \right]$ respectively, with $\mathcal{T} = \mu^2 / \sigma^2$. Note that a causal positive impact of past trends on future spreads should translate as a strong contribution to $C_\mu(\tau)$ for negative $\tau$. 
Interestingly, this is compatible with Fig.~\ref{fig:corr}(b), which confirms in a model-free fashion that the Zumbach-like coupling is important: past square trends increase future effective spread, or equivalently decrease future liquidity. While also slightly asymmetric, the volatility/spread correlation $C_\sigma(\tau)$ does not reveal such a level of asymmetry (see Fig. \ref{fig:corr}(c)). Fig.~\ref{fig:corr}(d) shows an even more pronounced asymmetry when we rescale the trend by the local volatility: $\mathcal{T}$ is a proxy of the autocorrelation of returns, independently of their amplitude. In this sense, it is a better signature of trend behaviour, as the volatility aspect of recent price changes is discarded. 

\section{Conclusion}\label{sec:conclusion}

In this work, we have proposed several actionable procedures to calibrate general Quadratic Hawkes models for order book events (market orders, limit orders, cancellations). One of the main features of such models is to encode not only the influence of past events on future events but also, crucially, the influence of past {\it price changes} on such events. We have shown that the empirically calibrated quadratic kernel (describing the part of the feedback that is independent of the sign of past returns) is well described by the shape postulated in \cite{blanc2017quadratic, fosset2019endogenous, dandapani2019quadratic}, namely:
\begin{itemize}\itemsep-0.3em 
    \item a diagonal contribution that captures past realised volatility, and
    \item  a rank-one contribution that captures the effect of past trends.
\end{itemize}
The latter contribution can be interpreted as the microstructural origin of the Zumbach effect: past trends, independently of their sign, tend to reduce the liquidity present in the order book, and therefore increase future volatility. As we have shown in our companion paper \cite{fosset2019endogenous}, such coupling can in fact be strong enough to destabilise the order book and lead to liquidity crises. 

One of the perhaps unexpected result of our calibration is that the Zumbach kernel is found to be a power-law of time for the futures contracts studied here, and not an exponential as was found in \cite{blanc2017quadratic} for US stock prices. Hence, all Hawkes kernels in our study are found to be power-laws of time. Furthermore, as in many previous studies \cite{filimonov2012quantifying, hardiman2013critical, bacry2015hawkes}, the rate of truly exogenous events is found to be much smaller than the total event rate, typically $1/5$ when all kernels are truncated beyond 1000 seconds, and probably even smaller when longer lags are taken into account, due to the slow decay of the kernels. These two features suggest that the system is close to a critical point -- in the sense that stronger feedback kernels would lead to instabilities. In our setting, we have shown that the effective spread (which is a measure of the (il-)liquidity of the order book) has itself a power-law tailed distribution, which we see as a precursor of the famous ``inverse cubic'' power-law tails of the return distribution (in the present context, see e.g. \cite{chicheportiche2014fine, blanc2017quadratic}). Such a power-law is not compatible with the alternative ``activated'' scenario proposed in  \cite{fosset2019endogenous}, which would rather suggest a bimodal distribution with a hump at large effective spreads. Hence, we favour at this stage the scenario of markets poised close to a point of instability, although the detailed mechanisms that lead to such a fine tuning are still somewhat obscure. We note that the near-criticality has also been argued to be crucial to understand the ``rough'' nature of volatility \cite{jaisson2016rough, jusselin, dandapani2019quadratic}. We believe that understanding these mechanisms is probably one of the most intellectually challenging (and exciting) issue for microstructure theorists. 

\section*{Acknowledgments}

We thank Jonathan Donier, Iacopo Mastromatteo, Jos\'e Moran, Mehdi Tomas, Stephen Hardiman and Mathieu Rosenbaum for fruitful discussions.
This research was conducted within the \emph{Econophysics \& Complex Systems Research Chair}, under the aegis of the Fondation du Risque, the Fondation de l'Ecole polytechnique, the Ecole polytechnique and Capital Fund Management.


\small

\clearpage

\bibliographystyle{unsrt}
\bibliography{bibs}

\clearpage

\appendix

\section{Appendix: Estimation procedure} \label{appendix:estimation}

Here we show how to practically estimate the kernels presented in section~\ref{section:calib} from empirical data. 
First, we detail the empirical estimators for averages and covariances, then focus on the time grids used for estimation, and finally discuss the numerical discretisation of  Eqs.~\eqref{eq:gqhawkes_moments}.
\paragraph{Covariance estimators} 
We assume that we have a sample of events of type $i$ that happen at times $\left(T_n^i\right)_n$, with $i=P$ for the price process. Calling $T$ the total length of observation, the estimators of the average intensities read:
\begin{subeqnarray}
    \Lambda^i & \approx & \displaystyle \frac{N_T^i}{T} \\
    \Delta_k & \approx & \displaystyle \frac{1}{T} \sum_n \left(\Delta_{T_n^P}\right)^k.
\end{subeqnarray}
For the covariance estimators, we use a classical approach for asynchronous data. Denoting $\Delta t$, $\Delta x$ the  time steps associated with times $t$ and $x$, one has:
\begin{subeqnarray}\label{eq:covariances_estimator}
\chi_{NN}^{ij}(t) & \approx &  \displaystyle  \frac{1}{T \Delta t}\sum_{n,p} \mathds{1}_{\{ T_n^i - T_p^j \in \, \left[t - \Delta t/2, t + \Delta t/2 \right] \}} - \Lambda^i \Lambda^j
 \\
\chi_{NP}^{i}(t) & \approx &  \displaystyle \frac{1}{T \Delta t}\sum_{n}
\Delta_{T_p^P}\mathds{1}_{\{ T_n^i - T_p^P \in \, \left[t - \Delta t/2, t + \Delta t/2 \right] \}}
 \\
\chi_{NP^2}^{i}(t) & \approx &  \displaystyle \frac{1}{T \Delta t}\sum_{n,p} \left( \Delta_{T_p^P} \right)^2 \mathds{1}_{\{ T_n^i - T_p^P \in \, \left[t - \Delta t/2, t + \Delta t/2 \right] \}} - \Lambda^i \Delta_2
 \\
\chi_{NPP}^{i}(t, x)&  \approx &  \displaystyle \frac{1}{T^2 \Delta t \Delta x}\sum_{n,p,q} \Delta_{T_p^P} \Delta_{T_q^P} \mathds{1}_{\{ T_n^i - T_p^P \in \, \left[t - \Delta t/2, t + \Delta t/2 \right], \; T_n^i - T_q^P \in \, \left[x - \Delta x/2, x + \Delta x/2 \right] \}}\\
\chi_{P^2P^2}(t) & \approx &  \displaystyle \frac{1}{T \Delta t}\sum_{n,p}\left( \Delta_{T_n^P} \right)^2 \left( \Delta_{T_p^P} \right)^2 \mathds{1}_{\{ T_n^P - T_p^P \in \, \left[t - \Delta t/2, t + \Delta t/2 \right] \}} - \Delta_2^2 .
\end{subeqnarray}
Note that, as mentioned above, one can choose different time grids for the Hawkes and price contributions.
Symmetry properties of the covariances enable us to estimate them only for positive times:
\begin{itemize}
    \item $\chi_{NN}^{ij}(-t) = \chi_{NN}^{ji}(t) $
    \item $\chi_{NP}^{i}(t) = 0 $ and $\chi_{NP^2}^{i}(t) = 0 $ for $t < 0$
    \item $\chi_{NPP}^{i}(t,x) = 0 $ for $ \min (t,x) < 0$
    \item $\chi_{P^2P^2}(-t) = \chi_{P^2P^2}(t) $.
\end{itemize}{}
One can reasonably assume that the covariances are $\mathcal{C}^1$ except in zero.


\paragraph{Choice of time grids}
 A good choice of time grid to estimate the kernels is provided in \cite{bacry2016estimation}. Indeed, quadrature points in log-scale are well suited to accurately account for long range behaviour in the norm of the kernels. Consistently, it is advised to have time intervals increasing at the same rate as the grid of points we use.
 On the other hand, taking disjoined intervals $\left[t - \Delta t/2, t + \Delta t/2 \right]$  enables fast computations of the covariances. To enforce all of this, we compute the differences between the quadrature points, sort them and take the cumulative sum. This gives the disjoined time intervals suited for fast computations. Then, with linear interpolation, we obtain the final values on the quadrature points.

\paragraph{Discretisation}
Equations \eqref{eq:gqhawkes_moments} can be discretised in two different ways, using properties of the covariances and  time grids. To show how to approximate the integrals, we provide an example of discretisation of $\int_{\mathbb{R}^+} f(s) ds$ for an arbitrary function $f$ using the time grid $\left( t_n \right)$. The two possibilities are:
\begin{itemize}
    \item The quadrature technique: 
    $\int_{\mathbb{R}^+} f(s) ds \approx \sum_n f \left(t_n\right) w_n$.
    \item The piece-wise $\mathcal{C}^1$ approximation: $    \int_{\mathbb{R}^+} f(s) ds  \approx \sum_n \frac{t_{n+1} - t_n}{2} \left[f\left(t_n^+\right) + f\left(t_{n+1}^-\right) \right]  $, 
with $f(x^+) = \lim\limits_{\substack{y \to x \\ y > x}} f(y) $\vspace{-0.3cm}\\
and $f(x^-) = \lim\limits_{\substack{y \to x \\ y < x}} f(y) $. 
\end{itemize}{}
The first approximation is very efficient to compute $\Tr K$ or $||\boldsymbol{\phi} ||$ using $(t_n^h)$ and $(w_n^h)$. The second  handles very well the behavior around zero and can be useful to solve Eq.~\eqref{eq:cov_hawkes}.

\section{Additional plots and tables}\label{appendix:plots}

\begin{figure}[h!]
  \centering
  \includegraphics[width=\columnwidth]{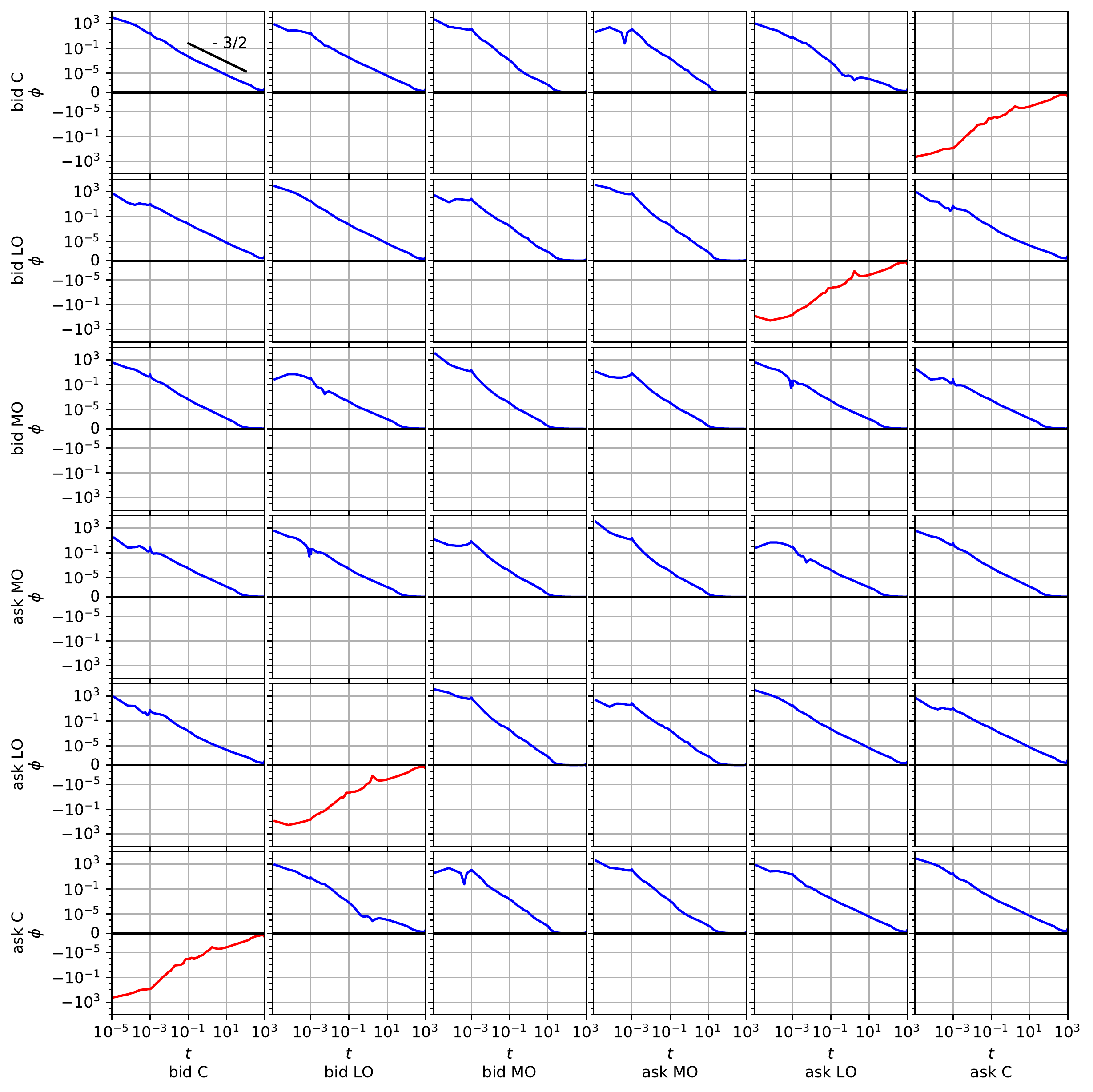}
  \caption{Hawkes kernels for the EURO STOXX futures contract between $2016/09/12$ and $2020/02/07$ ($t$ in seconds).}
  \label{fig:calibEurostock_hawkes_kernels}
\end{figure}

\vspace{6cm}

\pagebreak

\begin{figure}[h!]
  \centering
  \includegraphics[width=0.9\columnwidth]{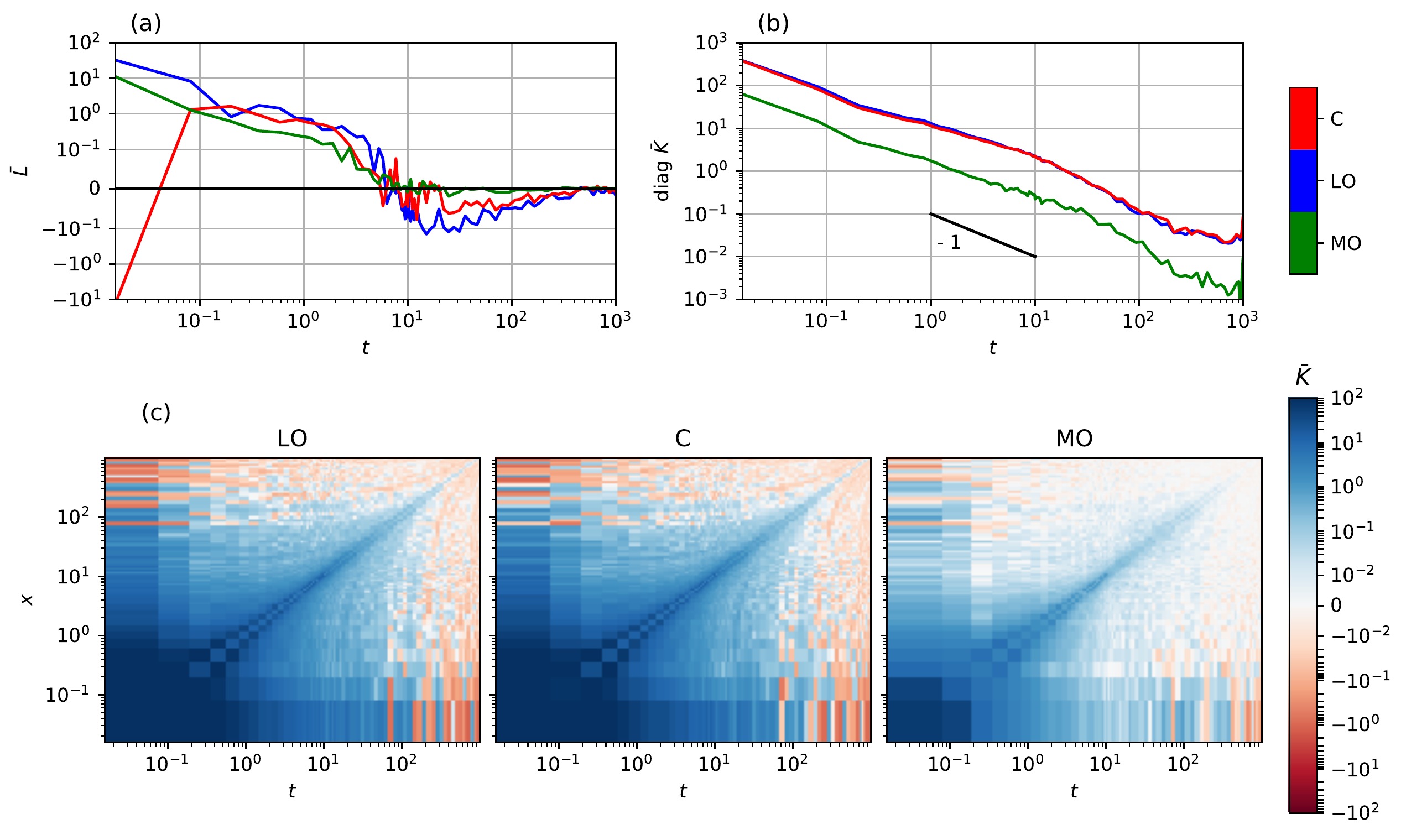}
  \caption{Raw effective kernels resulting from the calibration on the EURO STOXX futures contract between $2016/09/12$ and $2020/02/07$, without any smoothing procedure -- compare with Fig. \ref{fig:Eurostock_aggkernels}. (a) Linear kernels $\boldsymbol{ \bar L}$. (b) Diagonal of quadratic kernels $\boldsymbol{\bar K}_{\mathrm{d}}$.  (c) Full quadratic kernels $\boldsymbol{\bar K}(t,x).$}
  \label{fig:Eurostock_aggkernels_raw}
\end{figure}

\begin{figure}[h!]
  \centering
  \includegraphics[width=0.9\columnwidth]{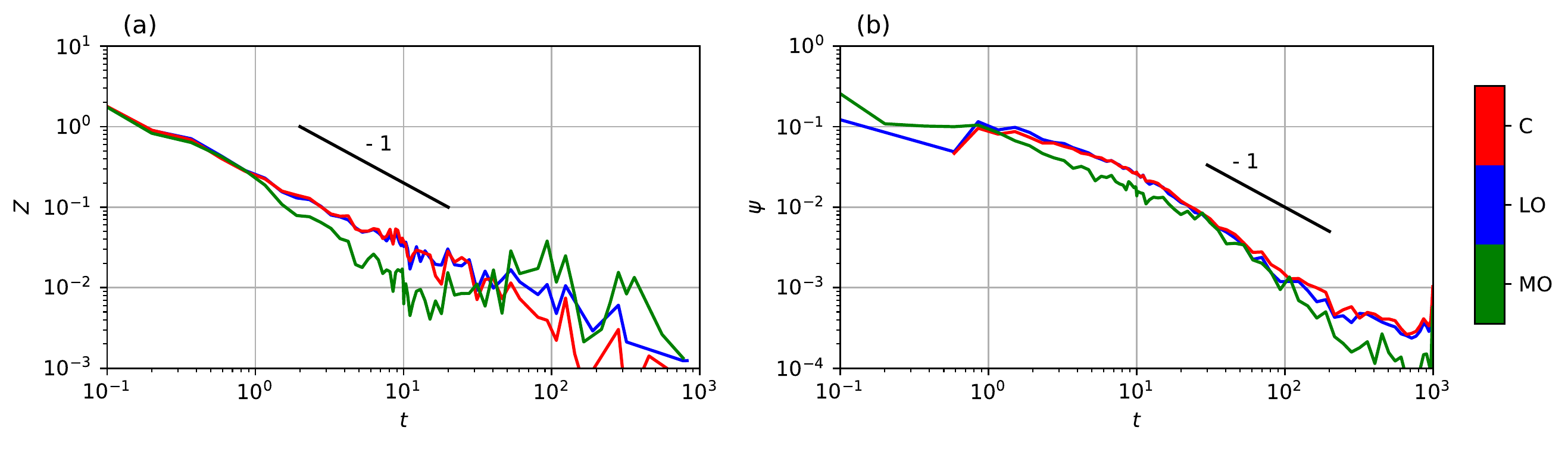}
  \caption{Zumbach approximation of the effective kernel $\bar{\boldsymbol{K}}$ on the EURO STOXX futures contract between $2016/09/12$ and $2020/02/07$ -- without any smoothing procedure -- compare with Fig. \ref{fig:zumbach_vol_kernels}. (a) Zumbach kernel $\boldsymbol Z$, (b) Volatility kernel $\boldsymbol \psi$.  Both kernels are normalised such that $||\boldsymbol \psi|| = || \boldsymbol Z^2 || =1$, with a cut-off in the time integrals at 1000 secs.}
  \label{fig:zumbach_vol_kernels_raw}
\end{figure}

\vspace{6cm}

\pagebreak

\begin{table}[b!]
\begin{centering}
\begin{tabular}{!{\vrule width 1pt}c!{\vrule width 1pt}c|c|c|c!{\vrule width 1pt}}
\noalign{\hrule height 1pt} 
 & &  \small $\mathrm{C}$ &  \small $ \mathrm{LO} $ &  \small $ \mathrm{MO} $  \\ \noalign{\hrule height 1pt}
\multirow{2}{*}{ \small $V \Tr \bar{K} \Delta_2 $} & \small EUROSTOXX  & \small $20.4$ & \small $ 18.8 $ & \small $2.1$ \\ \cline{2-5}
  & \small BUND & \small $5.7$ & \small $ 6.6 $ & \small $1.7$    \\ \noalign{\hrule height 1pt}
\multirow{2}{*}{ \small $V \bar{K}_1 \Delta_2 $} & \small EUROSTOXX  & \small $9.7$ & \small $ 8.6 $ & \small $0.5$  \\ \cline{2-5}
  & \small BUND & \small $3.5$ & \small $ 4.6 $ & \small $0.7$  \\ \noalign{\hrule height 1pt}
\multirow{2}{*}{ \small $V \bar{K}_{\mathrm{d}} \Delta_2 $} & \small EUROSTOXX  & \small $10.7$ & \small $ 10.1 $ & \small $1.6$  \\ \cline{2-5}
  & \small BUND & \small $2.2$ & \small $ 2.1$ & \small $1.0$  \\ \noalign{\hrule height 1pt}
\end{tabular}
\caption{Quadratic, Zumbach and volatility contributions to the liquidity rate of events (in shares per second).}
\label{tableQuadraticContributions}
\end{centering}
\end{table}

\begin{table}[t!]
\begin{centering}
\begin{tabular}{!{\vrule width 1pt}c!{\vrule width 1pt}c!{\vrule width 1pt}c|c|c!{\vrule width 1pt}}
\noalign{\hrule height 1pt} 
 & \small t (s) & \small  C & \small LO & \small MO  \\ \noalign{\hrule height 3pt}

\multirow{3}{*}{ \small $ \alpha_0^i / \Lambda^i $} & \small $10$  & \small $0.25$ & \small $ 0.14$ & \small $0.29$  \\ \cline{2-5}
& \small $ 100 $  & \small $0.24$ & \small $0.14$ & \small $0.27$ \\  \cline{2-5}
& \small $ 1000 $  & \small $0.23$ & \small $0.13$ & \small $0.27$ \\ \noalign{\hrule height 1pt}

\multirow{3}{*}{ \small $ \Delta_2 \Tr K^i / \Lambda^i $} & \small $10$  & \small $0.04$ & \small $ 0.03$ & \small $0.03$ \\ \cline{2-5}
& \small $ 100 $ & \small $0.05$ & \small $0.03$ & \small $0.04$  \\ \cline{2-5}
& \small $1000$ & \small $0.06$ & \small $0.04$ & \small $0.05$  \\ \noalign{\hrule height 1pt}

\multirow{3}{*}{ \small $ \Delta_2 \Tr \bar{K}^i / \Lambda^i $} & \small $10$  & \small $0.15$ & \small $ 0.16$ & \small $0.14$ \\ \cline{2-5}
& \small $ 100 $ & \small $0.23$ & \small $0.23$ & \small $0.21$ \\ \cline{2-5}
& \small $1000$ & \small $0.28$ & \small $0.28$ & \small $0.24$ \\ 
\noalign{\hrule height 3pt} 

\multirow{3}{*}{ \small $ \alpha_0^i / \sum_j ||\phi_{ij}|| \Lambda^j $} & \small $10$  & \small $0.35$ & \small $ 0.17$ & \small $0.42$  \\ \cline{2-5}
& \small $ 100 $  & \small $0.34$ & \small $0.17$ & \small $0.40$ \\  \cline{2-5}
& \small $ 1000 $  & \small $0.32$ & \small $0.16$ & \small $0.40$ \\ \noalign{\hrule height 1pt}

\multirow{3}{*}{ \small $ \Delta_2 K_1^i / \sum_j ||\phi_{ij}|| \Lambda^j $} & \small $10$  & \small $0.06$ & \small $ 0.05$ & \small $-0.01$  \\ \cline{2-5}
& \small $ 100 $  & \small $0.06$ & \small $ 0.05$ & \small $-0.01$ \\ \cline{2-5}
& \small $1000$ & \small $0.06$ & \small $ 0.05$ & \small $-0.01$ \\ \noalign{\hrule height 1pt}

\multirow{3}{*}{ \small $ \Delta_2 K_{\mathrm{d}}^i / \sum_j ||\phi_{ij}|| \Lambda^j $} & \small $10$  & \small  $-0.01$ &  \small $- 0.01$ & \small $0.06$ \\ \cline{2-5}
& \small $ 100 $  & \small $0.0$ & \small $-0.01$ & \small $0.08$ \\ \cline{2-5}
& \small $1000$ & \small $0.02$ & \small $0.0$ & \small $0.08$ \\ \noalign{\hrule height 3pt}

\multirow{3}{*}{ \small $\Delta_2 \bar{K}_1^i / \Lambda^i $} & \small $10$  & \small $0.13$ & \small $ 0.13$ & \small $0.05$ \\ \cline{2-5}
& \small $ 100 $  & \small $0.13$ & \small $0.13$ & \small $0.05$ \\ \cline{2-5}
& \small $1000$ & \small $0.13$ & \small $0.13$ & \small $0.05$  \\ \noalign{\hrule height 1pt}

\multirow{3}{*}{ \small $ \Delta_2 \bar{K}_{\mathrm{d}}^i / \Lambda^i $} & \small $10$  & \small  $0.02$ &  \small $ 0.03$ & \small $0.09$ \\ \cline{2-5}
& \small $ 100 $  & \small $0.10$ & \small $0.11$ & \small $0.16$  \\ \cline{2-5}
& \small $1000$ & \small $0.15$ & \small $0.15$ & \small $0.19$ \\ \noalign{\hrule height 1pt}
\end{tabular}
\caption{Different ratios between the quadratic contributions, base rates and  Hawkes contributions, truncated at different time scales $t(s)$ for the EURO STOXX futures contract between $2016/09/12$ and $2020/02/07$. The top three entries are the most important ones. For sake of simplicity, we have here approximated $\boldsymbol{K}_1$ as $ \left(1 - ||\boldsymbol{\phi}|| \right) \bar{\boldsymbol{K}}_1$ and $\boldsymbol{K}_{\mathrm{d}}$ as $ \left(1 - ||\boldsymbol{\phi}|| \right) \bar{\boldsymbol{K}}_{\mathrm{d}}$. }
\label{tableVolatility}
\end{centering}
\end{table}

\vspace{6cm}

\end{document}